\documentstyle[12pt,twoside,cosmion,epsf]{article}
\newcommand{\p}{\tilde{\phi}}
\newcommand{\x}{\tilde{\xi}}
\newcommand{\dep}{\delta \tilde{\phi}_k}
\newcommand{\dex}{\delta \tilde{\xi}_k}

\heads{Early Universe and Quantum Gravity}{Chaotic Reheating
\ \ {\rm D.I. Podolsky et al.}}

\begin{document}

\Arthead{1}{1}

\Title{Chaotic Reheating}
{D.I. Podolsky and A.A. Starobinsky}
{L.D. Landau Institute for Theoretical Physics}


\Abstract{We discuss the structure of parametric resonance which occurs
in the process of reheating after inflation with two interacting scalar 
fields. It is found that, for the case of a not too large coupling constant,
a quasi-homogeneous part of the second, initially subdominant scalar 
field may not be neglected due to its stochastic growth during 
inflation. This fact has strong consequences for the reheating stage: 
dynamics of background fields becomes chaotic after inflation --
it consists of subsequent chaotic and regular eras, and a tachyonic 
instability for inhomogeneous perturbations arises in the 
quantum reheating problem. This instability may pose a problem
for the standard reheating scenario. In order to avoid it, the coupling
constant should be either sufficiently small, or very large.}


\section{Introduction}

The simplest versions of the inflationary scenario of the early Universe,
which produced definite predictions about the total density of matter
and the power spectrum of adiabatic perturbations in the present-day 
Universe (later verified by observations), are based on the hypothesis
that our Universe had passed through a period of quasi-exponential 
expansion (the de Sitter, or inflationary stage) about 14 billion years in
the past, before the hot radiation-dominated (RD) Friedmann-Robertson-Walker 
(FRW) era occurred. This inflationary stage was driven by (at least) one 
effective scalar field (named inflaton) which was evolving sufficiently 
slow during inflation. Quantum vacuum fluctuations of this field induced
by space-time curvature during the de Sitter stage are the source of 
spatial inhomogeneities of space-time metric and matter density which
have led to formation of all structures in the present Universe including 
compact objects.  

However, to provide required smallness of initial perturbations, an
inflaton field should be very weakly coupled to other quantum fields
including "usual" ones (i.e., those of the standard model of strong, 
electromagnetic and weak interactions). Also, immediately after the end
of an inflationary stage, all energy density in the Universe is concentrated
in the quasi-homogeneous "cold" inflaton field $\phi$. So, to make matching 
with the subsequent hot RD stage possible, it is necessary, first, to
transmit the energy density to other quantum fields of matter 
and, second, to heat matter in the Universe.    
To achieve the first aim, the inflaton should be somehow coupled to other 
quantum fields. The simplest way to model a coupling to bosonic fields
of "usual" matter is to introduce a second scalar field $\xi$ interacting 
with the inflaton $\phi$. Of course, fermion fields and boson fields of a 
higher spin have to be added, too. Further, we assume the quartic 
interaction  $V_{int}= \frac{g^2}{2}\phi^2 \xi^2$.

In the perturbative regime, when occupation numbers of inhomogeneous modes
of all fields are small, energy transfer from the inflaton to other 
fields occurs through the process $\phi\phi\to\xi\xi$ which is rather slow.
The same process, as well the process $\phi\phi\to\phi\phi$ (in the case
of the $\frac{\lambda \phi^4}{4}$ inflaton self-interaction) and other
non-linear processes, leads to reheating of matter, too. However, it was
found in 90th that sometimes the energy transfer and reheating may proceed 
much quicker if they occur in a regime of the broad parametric resonance, 
see \cite{p1}$^{-}$\cite{preheat} (a narrow parametric resonance is usually 
not effective due to expansion of the Universe). The structure of the 
resonance is rather complicated \cite{preheat}: there is a 
stability - instability chart in the variables $\left(\frac{g^2}{\lambda}, 
k \right)$ where $k$ is the inverse wave-length of a $\xi$ - mode.

In the present work, we will show that this picture actually takes place 
for $\kappa =\frac{g^2}{\lambda} \gg 1$ only. As we will see, in the 
opposite case the conjecture of a single field driven inflation is not 
correct. This results in a drastic change of dynamics during a reheating 
stage. Dynamical chaos arises in the quasi-homogeneous background problem, 
that is a general feature for Hamiltonian systems with a large number of 
degrees of freedom (see, e.g., \cite{ch1}$^{,}$\cite{ch2} regarding chaotic 
dynamics in isotropic cosmology). 
There is even more dramatic change in the quantum inhomogeneous reheating 
problem: a tachyonic instability arises for sufficiently large $\kappa$ and 
long wavelengths. It may present a problem for the standard reheating model
because of generation of strong inhomogeneities. To exclude a tachyonic 
instability without qualitative change of the theory, one need the coupling 
constant $\kappa$ to be small as compared with unity.

\section{Inflationary stage}

Our model of the isotropic flat Universe is similar to that introduced in 
\cite{preheat}:
\begin{equation}
\ddot{\phi} + 3H\dot{\phi} + \lambda \phi^3 + g^2 \xi^2 \phi = 0~,
\label{phif}
\end{equation}
\begin{equation}
\ddot{\xi} + 3H\dot{\xi} + g^2 \phi^2 \xi = 0~,
\label{xif}
\end{equation}
\begin{equation}
H^2 = \frac{8 \pi}{3 M_P^2} \left( \frac{1}{2}{\dot{\phi}}^2 +
\frac{1}{2}{\dot{\xi}}^2 + \frac{\lambda}{4} \phi^4 +
 \frac{g^2}{2}\phi^2 \xi^2 \right)~.
\label{Hf}
\end{equation}

The usual point of view \cite{preheat} is that the inflationary stage is driven 
by the field $\phi$ and the dynamics of field $\xi$ is not important until the 
inflation is finished. Actually a criterion of validity of this statement is 
given by a value of the effective mass of the field $\xi$. If initial 
conditions for (\ref{phif} - \ref{Hf}) are such that the mass of $\xi$ is much 
larger than the Hubble parameter $H$, inflation becomes single-field driven 
soon after its beginning, and the scenario described in~\cite{preheat} is 
realized. In fact, this condition limits the coupling constant $\kappa = 
g^2 / \lambda$: if the inequality $m_\xi^2 = g^2 \phi^2 \gg H^2 =
2\pi G \lambda \phi^4 /3$ remains valid until the end of inflation (when 
$G \phi^2 \sim 1$), then $\kappa \gg 1$ ($M_P^2=G^{-1}$).

What happens when the coupling constant $\kappa$ is not large? Let us suppose 
that $\xi$ is never dynamically important in this regime, too:
\begin{equation}
g^2 \xi^2 \ll \lambda \phi^2~.
\label{in1}
\end{equation}
Here $\phi$ and $\xi$ may be understood both as operators and stochastic 
quasi-classical quantities. Furthermore, let us consider the period of inflation 
when fluctuations in $\phi$ may be neglected, so that the background field $\phi$ 
may be considered as a deterministic quantity. Its evolution as a function of 
$\tau = \ln(a(t)/a_f)$, where $a_f$ is the value of the scale factor at the end 
of inflation (thus, $\tau < 0$), is given by the equation
\begin{equation}
\phi^2 = - \frac{\tau}{\pi G},~~~|\tau | \gg 1~.
\label{phiev}
\end{equation}
Then the evolution of $\xi$ may be calculated in the Gaussian approximation 
(i.e., assuming that its distribution is Gaussian). 
The following inequality is valid in the regime under consideration:
$m_\xi^2 = g^2 \phi^2 \ll H^2 = 2\pi G \lambda \phi^4 /3$. Since $G \phi^2 
\gg 1$ during inflation, it is satisfied, if $g^2$ is not too much larger than 
$\lambda$ (in fact, this regime takes place everywhere on the "phase diagram" 
of our theory except for the region $\kappa \gg 1$!). The equation for the 
evolution of $z \equiv \pi G \langle \xi^2 \rangle$ reads:
\begin{equation}
\frac{dz}{d\tau} = - \frac{2 m_\xi^2 z}{3 H^2} + \frac{GH^2}{4 \pi} =
\frac{g^2 z}{\lambda \tau} + \frac{\lambda \tau^2}{6 \pi^2}.
\label{zdyn}
\end{equation}

Let us take $z = 0$ for $|\tau | = \tau_0$ as an initial condition. Then
\begin{equation}
z = C |\tau |^{g^2 / \lambda} - \frac{\lambda |\tau |^3}{6\pi^2
\left( 3 - \frac{g^2}{\lambda}   \right)},~~~C =
\frac{\lambda \tau_0 ^{3 - \frac{g^2}{\lambda}}} {6 \pi^2 \left( 3 - 
\frac{g^2}{\lambda}\right)}~.
\label{zsol}
\end{equation}

As we can see, there is a stochastic growth of $\langle \xi^2 \rangle$ due
to the second term on the right-hand side. Thus, we come to an inconsistency 
with the conjecture (\ref{in1}): it is impossible to neglect the influence
of $\xi$ on the large-scale inflationary dynamics, and one needs to solve 
a two-field problem if $\kappa$ is not large as compared with unity. This fact 
results in a radical change of reheating dynamics, too. Now we are going to 
discuss this question in detail.

\section{Reheating: solution of a homogeneous problem}

It is convenient \cite{preheat} to introduce new "conformal" variables 
$\p = a\phi$, $\x = a\xi$, $\eta = \int \frac{dt}{a}$ soon after the end 
of inflation. The equations (\ref{phif} - \ref{Hf}) take the following form
in these variables (a prime denotes the derivative with respect to the 
conformal time $\eta$):
\begin{equation}
\p '' - \frac{a''}{a}\p + \lambda \p^3 + g^2 \x^2 \p = 0~,
\label{pconff}
\end{equation}
\begin{equation}
\x '' - \frac{a''}{a}\x  + g^2 \p^2 \x = 0~,
\label{xconff}
\end{equation}
\begin{equation}
\left( a' \right)^2 = \frac{8\pi}{3M_P^2}
\left( \frac{1}{2}\left( \p' - \frac{a'}{a}\p \right)^2
+ \frac{1}{2}\left( \x' - \frac{a'}{a}\x \right)^2  +
 \frac{\lambda}{4} \p^4 + \frac{g^2}{2}\p^2 \x^2  \right)~.
\label{aconff}
\end{equation}
After the end of the inflationary stage, terms containing $\frac{a''}{a}$ 
and $\frac{a'}{a}$ may be neglected. Then the system
(\ref{pconff}-\ref{aconff}) is drastically simplified making the problem
almost Hamiltonian:
\begin{equation}
\p '' + \lambda \p^3 + g^2 \x^2 \p = 0~,
\label{pconf}
\end{equation}
\begin{equation}
\x ''  + g^2 \p^2 \x = 0~,
\label{xconf}
\end{equation}
\begin{equation}
\left( a' \right)^2 = \frac{8\pi}{3M_P^2}
\left( \frac{1}{2}\left( \p' \right)^2
+  \frac{1}{2}\left( \x' \right)^2  +
\frac{\lambda}{4} \p^4 +
  \frac{g^2}{2}\p^2 \x^2  \right)  =
\frac{8\pi}{3M_P^2} E~.
\label{aconf}
\end{equation}

Structure of its solutions can easily be understood using the projection of 
its phase space onto the plain $\p' = 0$, $\x' = 0$. The region where the 
system moves is bounded by the curve
\begin{equation}
\x = \pm \frac{1}{g|\p |}~\sqrt{2\left(E - \frac{\lambda}{4}\p^4 
\right)}~.
\label{boundary}
\end{equation}

There are two essentially different regimes of the system behaviour. The 
first is realized when $\x$ and $\p$ are of the same order. Then both
fields oscillate. The system moves somewhere in the region near the origin, 
and its motion is substantially chaotic because of chaotic character of 
energy redistribution between $\p$ and $\x$ degrees of freedom. For larger
$\kappa$, chaos becomes stronger. We call these periods "chaotic eras".

The second regime is realized when a strong fluctuation of 
$\left( \x' \right)^2$ occurs, and the system leaves the central region 
percolating into a "tube" under or below this region. This tube is narrow, 
so the system cannot live inside it during an infinite time. The lower the 
energy of the system $E$ and the higher the control parameter $\kappa$ are, 
the larger is the number of oscillations inside the tube. We call
these periods "regular" (because of the possibility of finding an approximate 
solution for $\p$ and $\x$) or "double - period eras" (because of the 
analogy with intermittency in hydrodynamics).

\section{Special solution with synchronous oscillations}

Let us consider a particular case when initial conditions are chosen in 
such a way that homogeneous components of $\p$ and $\x$ are proportional to 
each other at the end of inflation. Of course, generically it is not so. 
However, we will see below that this special solution correctly reproduces 
almost all features characteristic for a generic case.

Examining the equations (\ref{pconf},\ref{xconf}) it is easy to find that if
$\p = \alpha \x$, then the constant $\alpha$ is fixed:
\begin{equation}
\label{alphatoy}
\alpha^2 = \frac{g^2}{g^2 - \lambda} = \frac{\kappa}{\kappa - 1},
\end{equation}
and we have the following equation for $\p$:
\begin{equation}
\label{prealtoy}
\p '' + g^2 \p^3 = 0
\end{equation}
Of course, there is no chaos for this solution. Now we consider fluctuations 
of $\p$ and $\x$ on this background. Introducing the Fourier mode 
decomposition as usual, we get:
\begin{equation}
\label{dxik}
{\dex}'' + \left(k^2 + g^2\p^2\right)\dex + 2g^2 \x \p \dep = 0~,
\end{equation}
\begin{equation}
\label{dphik}
{\dep}'' + \left( k^2 + 3\lambda \p^2 + g^2\x^2\right)\dep + 
2g^2 \x \p \dex = 0~.
\end{equation}
When $\dep$ and $\dex$ modes are strongly coupled to each other, it is not 
clear how to determine the number of particles as well as Floquet indices. 
Nevertheless, it is possible to introduce new decoupled perturbations:
\begin{equation}
\label{Sig}
\Sigma_k = A \dex + B \dep ,
\end{equation}
\begin{equation}
\label{Del}
\Delta_k = C \dex + D \dep 
\end{equation}
where $A, B, C, D$ are some constants to be determined. 
Omitting a simple calculation, we present final equations for these rotated 
perturbations:
\begin{equation}
\label{sigeq}
\Sigma_k '' + (k^2 + 3g^2\phi^2)\Sigma_k = 0,
\end{equation}
\begin{equation}
\label{deleq}
\Delta_k '' + (k^2 + \frac{2-\kappa}{\kappa} g^2\phi^2)\Delta_k = 0~.
\end{equation}

Therefore, first, the presence of $\xi\not= 0$ 
results in rotation in the space of field perturbations. Now correct 
variables are the rotated ones. It should be noted that the definition of 
fields $\Sigma_k$, $\Delta_k$ strongly depends on $\kappa$, so there is no 
hope to construct something like a stability-instability chart for 
arbitrary initial fields $\p$ and $\x$. 

Second, the field $\Sigma_k$ can be interpreted as an "inflaton" with the 
wave vector $k$, because the structure of (\ref{sigeq}) is the same as 
for the inflaton mode if $\x = 0$.

Third, there is a tachyonic instability for $\Delta_k$ when $\kappa > 2$.
It means that there is an avalanche-like creation of 
excitations for this mode, but these excitations are quite unlike usual
particles. Instead, we have a chaotic growth of spatial inhomogeneities. 
It is interesting to note that this 
special toy solution gives an opportunity to investigate the system 
behaviour near the boundary of chaotic instability. The boundary for the 
instability for this toy solution is sharp -- there is no instability for 
all modes $k$ if $\kappa \le 2$. On the other hand, $\kappa > 1$ is 
necessary for this solution to exist at all.

It turns out that almost all these consequences are valid in a generic 
case (background fields $\p$ and $\x$ are both nonzero and independent) 
if we consider the WKB regime of an inhomogeneous problem. There is only 
one difference -- it is hardly possible to determine an exact boundary 
of instability in this case. This will be discussed elsewhere.

\section{Discussion}

We see that the structure of resonance in the standard reheating theory 
(\ref{phif} - \ref{Hf}) can be more complicated than the picture 
presented in \cite{preheat}. The stability - instability chart exists 
for $\kappa \gg 1$ only (i.e., in the broad resonance regime), and there 
is no possibility of its constructing for not too large $\kappa$ because of 
the appearance of chaos in the homogeneous background problem and the 
necessity to rotate modes. The underlying reason is the stochastic growth 
of $\langle \xi^2 \rangle$ during inflation. As a result, 
quasi-homogeneous backgrounds for both scalar fields $\phi$ and $\xi$ are 
non-zero by the end and after inflation. Thus, chaos occurs in the 
homogeneous reheating 
problem, which is a usual situation for Hamiltonian system with a 
sufficiently large dimension of the phase space. Note that the 
Toda-Brumer necessary criterion for the global dynamical chaos (a 
negative determinant of the matrix $||\partial^2 U(\phi , \xi)/
\partial\phi\,\partial\xi||$) is fulfilled for our model. Generically, 
chaotic mixing in the $(\p ,\x)$ plane rapidly provides equipartition of 
energy density between the inflaton ($\phi$) and the other matter field 
($\xi$). Thus, the first aim of reheating is achieved. 

Evolution of conformally transformed fields $\p$ and $\x$ consists of
consequent chaotic ($|\p|\sim |\x|$) and regular ($|\p|\ll |\x|$) eras. 
This phenomenon is similar to intermittency in the theory of nonlinear 
dynamics. 
There, it is well known that the number of chaotic eras grows while 
increasing an external parameter of a system ($1/\kappa$ in our case).
This analogy leads to a conjecture that a critical $\kappa_c$ exists such 
that regular eras disappear for $\kappa < \kappa_c$. 

Dynamics of quantum reheating in the case of nonzero $\xi (t)$ differs from 
the picture \cite{preheat} even without dynamical chaos in the problem
(\ref{phif}-\ref{Hf}). The first new feature is a time-dependent 
rotation in the space of modes $\dep$ and $\dex$. This effect is not 
unexpected: while our work was in preparation, papers discussing the same 
effect have appeared \cite{nilles}$^,$\cite{rot}. Its necessity is clear 
from the physical point of view: the Universe should be filled up with 
particle-like excitations after a reheating stage. It is well known that 
the term "particle" is correctly defined for an oscillator with a variable 
frequency in the WKB-regime only. The correct WKB-solution may not be 
constructed without mode rotation in our case.  

The second, more important, feature is the appearance of tachyonic 
instability for sufficiently large $\kappa$ and long wavelengths. This 
effect is different from that recently found in the theory of reheating in 
the course of spontaneous symmetry breaking after the end of the hybrid 
inflation \cite{tachyon}$^,$\cite{tach1}. The latter effect occurs in  
case the oscillation frequency of background fields immediately after 
inflation is much more than the Hubble parameter $H$ at this moment. In our 
model, this is not the case and there is no spontaneous symmetry breaking. 
So, initially it was hardly possible to expect a vacuum instability with 
respect to the creation of tachyonic excitations. 

The tachyonic instability in our model is more like an instability in
chaotic dynamical systems with a positive largest Lyapunov exponent 
\footnote {Regarding long wavelength perturbations, a close point of view was 
expressed in the recent paper \cite{zib} which appeared when our paper was 
prepared for publication.}. The question arises immediately: is it a 
desirable or undesirable property? On one hand, this instability is closely 
connected with chaotic mixing leading to a rapid transfer of energy 
density from the inflaton to other fields. On the other hand, it produces 
strong spatial inhomogeneities in the Universe (and inhomogeneously 
distributed high-energy particles as was in the case of the broad 
parametric resonance type preheating \cite{p1}$^,$\cite{preheat}). Special 
investigation is needed to track the further fate of these inhomogeneities. 
At this moment, we want to state only that, to avoid a strong tachyonic 
instability, the coupling constant $\kappa$ 
should be either sufficiently small, or very large (in the latter case, 
there is no chaos in the homogeneous reheating problem, and the results of 
\cite{preheat} remain valid).

This research was partially supported by the Russian Foundation for
Basic Research, grants 99-02-16224 and 00-15-96699, and by the Research
Programme "Quantum Macrophysics" of the Russian Academy of Sciences.



\begin{thebibliography}{99} \itemsep=-5pt
\bibitem{p1} Kofman, L.A., Linde, A.D. and Starobinsky, A.A. 1994, Phys. 
Rev. Lett. {\bf 73}, 3195.
\bibitem{p2} Shtanov, Y., Traschen, J. and Brandenberger, R. 1995, Phys. 
Rev. D {\bf 51}, 5438.
\bibitem{p3} Boyanovsky, D., de Vega, H.J., Holman, R., Lee, D.S. and 
Singh, A. 1995, Phys. Rev. D {\bf 51}, 4419.
\bibitem{p5} Khlebnikov, S. and Tkachev, I. 1996, Phys. Rev. Lett. {\bf 77}, 
219. 
\bibitem{p6} Riotto, A. and Tkachev, I. 1996, Phys. Lett. {\bf B385}, 57. 
\bibitem{p7} Boyanovsky, D., de Vega, H.J., Holman, R., Lee, D.S., Singh, A. 
and Salgado, J.F.J. 1996, Phys. Rev. D {\bf 54}, 7570.
\bibitem{p8} Kofman, L.A., Linde, A.D. and Starobinsky, A.A. 1997, Phys. 
Rev. D {\bf 56}, 3258. 
\bibitem{preheat} Green, P.B., Kofman, L.A., Linde, A.D. and Starobinsky, 
A.A. 1997, Phys. Rev. D {\bf 56}, 6175.
\bibitem{ch1} Cornish, N.J. and Levin, J.J. 1996, Phys. Rev. D {\bf 53}, 
3022. 
\bibitem{ch2} Easther, R. and Maeda, K. 1999, Class. Quant. Grav. {\bf 16}, 
1637. 
\bibitem{nilles} Nilles, H.P., Peloso, M. and Sorbo, L. 2001, JHEP 
{\bf 104}, 4.
\bibitem{rot} Cormier, D., Heitmann, K. and Mazumdar, A. 2001, preprint 
hep-ph/0105236.
\bibitem{tachyon} Felder, G., Garcia-Bellido, J., Green, P.B., Kofman, L., 
Linde, A. and Tkachev, I. 2001, Phys. Rev. Lett. {\bf 87}, 011601.
\bibitem{tach1} Felder, G., Kofman, L. and Linde A. 2001, Phys. Rev. D
{\bf 64}, 123517.                                         
\bibitem{zib} Zibin, J.P. 2001, preprint astro-ph/0108008.

\end{thebibliography}
\end{document}